# Relativistic symmetries in Rosen-Morse potential and tensor interaction using the Nikiforov-Uvarov method

Sameer M. Ikhdair[1*], Majid Hamzavi[2**]

[1]*Physics Department, Near East University, Nicosia, North Cyprus, Mersin 10, Turkey*

[2]*Department of Basic Sciences, Shahrood Branch, Islamic Azad University, Shahrood, Iran*

[*] Corresponding Author sikhdair@neu.edu.tr

Tel: +90-392-2236624; Fax: +90-392-2236622

[**] majid.hamzavi@gmail.com

Tel.:+98 273 3395270, Fax: +98 273 3395270

**Abstract**

Approximate analytical bound-state solutions of the Dirac particle in the field of both attractive and repulsive RM potentials including Coulomb-like tensor (CLT) potential are obtained for arbitrary spin-orbit quantum number $\kappa$. The Pekeris approximation is used to deal with the spin-orbit coupling terms $\kappa(\kappa\pm1)r^{-2}$. In the presence of exact spin and pseudospin (p-spin) symmetries, the energy eigenvalues and the corresponding normalized two-component wave functions are found by using the parametric generalization of the Nikiforov-Uvarov (NU) method. The numerical results show that the CLT interaction removes degeneracies between spin and p-spin state doublets.

**Keywords:** Dirac equation, RM potential, CLT potential, spin and p-spin symmetries, NU method, approximation schemes

**PACS:** 03.65.Ge, 03.65.Fd, 03.65.Pm, 02.30.Gp

## 1. Introduction

In the framework of the Dirac equation, the spin symmetry occurs when the difference of the scalar $S(\vec{r})$ and vector $V(\vec{r})$ potentials is a constant, i.e., $\Delta(r)=C_s$ and the p-spin symmetry occurs when the sum of the scalar and vector potentials is a constant, i.e., $\Sigma(r)=C_{ps}$ [1-3]. The spin symmetry is relevant for mesons [4]. The p-spin symmetry concept has been applied to many systems in nuclear physics and related areas [2-7] and used to explain features of deformed nuclei [8], the super-deformation [9] and to establish an effective nuclear shell-model scheme [5,6,10]. The pspin symmetry introduced in nuclear theory refers to a quasi-degeneracy of the single-nucleon doublets and can be characterized with the non-relativistic quantum numbers ($n, l, j = l + 1/2$) and ($n-1, l+2, j = l+3/2$), where $n$, $l$ and $j$ are the single-nucleon radial, orbital and total angular momentum quantum numbers for a single particle, respectively [5,6]. The total angular momentum is given as $j = \tilde{l} + \tilde{s}$, where $\tilde{l} = l + 1$ is a p-angular momentum and $\tilde{s} = 1/2$ is a p-spin angular momentum. In real nuclei, the p-spin symmetry is only an approximation and the quality of approximation depends on the p-centrifugal potential and p-spin orbital potential [11]. Alhaidari *et al.* [12] investigated in detail the physical interpretation on the three-dimensional Dirac equation in the context of spin symmetry limitation $\Delta(r) = 0$ and p-spin symmetry limitation $\Sigma(r) = 0$.

The exact solutions of the Dirac equation for the exponential-type potentials are possible only for the $s$-wave ($l = 0$ case). However, for $l$-states an approximation scheme has to be used to deal with the centrifugal and p-centrifugal terms. Many authors have used different methods to study the partially exactly solvable and exactly solvable Schrödinger, Klein-Gordon (KG), Dirac and semi-relativistic Salpeter equations in $1D$, $3D$ and arbitrary $D$-dimension for different potentials [13-25]. In the context of spatially-dependent mass, we have also used and applied a proposed approximation scheme [26] for the centrifugal term to find a quasi-exact analytic bound-state solution of the radial KG equation with spatially-dependent effective mass for scalar and vector Hulthén potentials in any arbitrary dimension $D$ and orbital angular momentum quantum number $l$ within the framework of the NU method [26-28].

Another physical potential is the Rosen-Morse potential [29] expressed in the form

$$V(r) = -V_1 \operatorname{sech}^2 \alpha r + V_2 \tanh \alpha r, \tag{1}$$

where $V_1$ and $V_2$ denote the depth of the potential and $\alpha$ is the range of the potential. The RM potential is useful for describing interatomic interaction of the linear molecules and helpful for discussing polyatomic vibration energies such as the vibration states of $NH_3$ molecule [29]. It is shown that the RM potential and its PT-symmetric version are the special cases of the five-parameter exponential-type potential model [30,31]. The exact energy spectrum of the trigonometric RM potential has been investigated by using supersymmetric (SUSY) and improved quantization rule methods [32,33].

Lisboa *et al.* [34] have studied a generalized relativistic harmonic oscillator for spin-$1/2$ fermions by solving Dirac equation with quadratic vector and scalar potentials including a linear tensor potential with spin and p-spin symmetry. Further, Akçay [35] has shown that the Dirac equation for scalar and vector quadratic potentials and Coulomb-like tensor potential with spin and p-spin symmetry can be solved exactly. In these works, it has been found out that the tensor interaction removes the degeneracy between two states in the p-spin doublets. The tensor coupling under the spin and p-spin symmetry has also been studied in [36,37]. Further, the nuclear properties have been studied by using tensor couplings [38,39]. Very recently, various types of potentials like Hulthén [40] and Woods-Saxon [41] including Coulomb-like potential have been studied with the conditions of spin and p-spin symmetry. The p-spin symmetric solution of the Dirac-Morse problem with the Coulomb-like tensor potential for any spin-orbit quantum number $\kappa$ has been studied [42].

In the present work, our aim is to present the analytical solutions of the Dirac-RM problem with the Coulomb-like tensor potential for arbitrary spin-orbit quantum numbers $\kappa$ that are not large and vibrations of the small amplitude about the minimum. This approximation has been introduced by Lu [43] and used in [44] to deal with the centrifugal term near the minimum point $r = r_e$.

The paper is structured as follows: In Sect. 2, we outline the NU method. In Sec. 3, we obtain the analytical spin and p-spin symmetric bound state solutions of the (3+1)-dimensional Dirac equation for the RM potential model including CLT potential by means of the NU method. Finally, the relevant conclusion is given in Sect. 4.

## 2. Parametric Generalization of the NU method

The NU method is used to solve second order differential equations with an appropriate coordinate transformation $s = s(r)$ [27]

$$\psi_n''(s)+\frac{\tilde{\tau}(s)}{\sigma(s)}\psi_n'(s)+\frac{\tilde{\sigma}(s)}{\sigma^2(s)}\psi_n(s)=0\,, \tag{2}$$

where $\sigma(s)$ and $\tilde{\sigma}(s)$ are polynomials, at most of second degree, and $\tilde{\tau}(s)$ is a first-degree polynomial. To make the application of the NU method simpler and direct without need to check the validity of solution. We present a shortcut for the method. So, at first we write the general form of the Schrödinger-like equation (2) in a more general form applicable to any potential as follows [45]

$$\psi_n''(s)+\left(\frac{c_1-c_2s}{s(c_3-c_4s)}\right)\psi_n'(s)+\left(\frac{-As^2+Bs-C}{s^2(c_3-c_4s)^2}\right)\psi_n(s)=0, \tag{3}$$

satisfying the wave functions

$$\psi_n(s)=\phi(s)y_n(s). \tag{4}$$

Comparing (3) with its counterpart (2), we obtain the following identifications:

$$\tilde{\tau}(s)=c_1-c_2s,\quad \sigma(s)=s(c_3-c_4s),\quad \tilde{\sigma}(s)=-As^2+Bs-C, \tag{5}$$

where $c_i$ $(i=1,2,3,4)$, $A$, $B$ and $C$ are constant coefficients to be found for the potential model. Following the NU method [27], we obtain the followings [45],

(i) the relevant constant:

$$c_5=\frac{1}{2}(c_3-c_1),\qquad c_6=\frac{1}{2}(c_2-2c_4),$$

$$c_7=c_6^2+A,\qquad c_8=2c_5c_6-B,$$

$$c_9=c_5^2+C,\qquad c_{10}=c_4(c_3c_8+c_4c_9)+c_3^2c_7,$$

$$c_{11}=\frac{2}{c_3}\sqrt{c_9},\ c_3\neq 0,\qquad c_{12}=\frac{2}{c_3c_4}\sqrt{c_{10}},\ c_4\neq 0,$$

$$c_{13}=\frac{1}{c_3}\left(c_5+\sqrt{c_9}\right),\qquad c_{14}=\frac{1}{c_3c_4}\left(\sqrt{c_{10}}-c_4c_5-c_3c_6\right),$$

$$c_{15}=\frac{2}{c_3}\sqrt{c_{10}},\qquad c_{16}=\frac{1}{c_3}\left(\sqrt{c_{10}}-c_4c_5-c_3c_6\right), \tag{6}$$

(ii) the essential polynomial functions:

$$\pi(s)=c_5+\sqrt{c_9}-\frac{1}{c_3}\left(c_4\sqrt{c_9}+\sqrt{c_{10}}-c_3c_6\right)s, \tag{7}$$

$$k=-\frac{1}{c_3^2}\left(c_3c_8+2c_4c_9+2\sqrt{c_9c_{10}}\right), \tag{8}$$

$$\tau(s)=c_3+2\sqrt{c_9}-\frac{2}{c_3}\left(c_3c_4+c_4\sqrt{c_9}+\sqrt{c_{10}}\right)s, \tag{9}$$

$$\tau'(s)=-\frac{2}{c_3}\left(c_3c_4+c_4\sqrt{c_9}+\sqrt{c_{10}}\right)<0. \tag{10}$$

(iii) The energy equation:

$$c_2n-(2n+1)c_6+\frac{1}{c_3}(2n+1)\left(\sqrt{c_{10}}+c_4\sqrt{c_9}\right)+n(n-1)c_4$$
$$+\frac{1}{c_3^2}\left(c_3c_8+2c_4c_9+2\sqrt{c_9c_{10}}\right)=0. \tag{11}$$

(iv) The wave functions

$$\rho(s)=s^{c_{11}}\left(c_3-c_4s\right)^{c_{12}}, \tag{12}$$

$$\phi(s)=s^{c_{13}}\left(c_3-c_4s\right)^{c_{14}},\ c_{13}>0,\ c_{14}>0, \tag{13}$$

$$y_n(s)=P_n^{(c_{11},c_{12})}\left(c_3-2c_4s\right),\ c_{11}>-1,\ c_{12}>-1,\ s\in\left[(c_3-1)/2c_4,(c_3+1)/2c_4\right], \tag{14}$$

$$\psi_{n\kappa}(s)=N_{n\kappa}s^{c_{13}}\left(c_3-c_4s\right)^{c_{14}}P_n^{(c_{11},c_{12})}\left(c_3-2c_4s\right), \tag{15}$$

where $P_n^{(\mu,\nu)}(x),\ \mu>-1,\ \nu>-1$, and $x\in[-1,1]$ are Jacobi polynomials with

$$P_n^{(\alpha,\beta)}(1-2s)=\frac{(\alpha+1)_n}{n!}\,{}_2F_1(-n,1+\alpha+\beta+n;\alpha+1;s), \tag{16}$$

and $N_{n\kappa}$ is a normalization constant. When $c_4=0$, the Jacobi polynomial turn to be the generalized Laguerre polynomial and the constants relevant to this polynomial change are

$$\lim_{c_4\to 0}P_n^{(c_{11},c_{12})}\left(c_3-2c_4s\right)=L_n^{c_{11}}\left(c_{15}s\right),$$

$$\lim_{c_4\to 0}\left(c_3-c_4s\right)^{c_{14}}=\exp\left(-c_{16}s\right),$$

$$\psi_{n\kappa}(s)=N_{n\kappa}\exp\left(-c_{16}s\right)L_n^{c_{11}}\left(c_{15}s\right), \tag{17}$$

where $L_n^{c_{11}}\left(c_{15}s\right)$ are the generalized Laguerre polynomials and $N_{n\kappa}$ is a normalization constant.

## 3. Dirac bound states of the RM potential and CLT potential

The Dirac equation for a particle of mass $m$ moving in the field of attractive radial scalar $S(r)$, repulsive vector $V(r)$ and tensor $U(r)$ potentials (in the relativistic units $\hbar=c=1$) takes the form [34]

$$\left[\vec{\alpha}\cdot\vec{p}+\beta\left(m+S\left(r\right)\right)+V\left(r\right)-i\beta\vec{\alpha}\cdot\hat{r}U(r)\right]\psi\left(\vec{r}\right)=E\psi\left(\vec{r}\right), \tag{18}$$

with $E$ is the relativistic energy of the system, $\vec{p}=-i\vec{\nabla}$ is the three-dimensional (3D) momentum operator and $\vec{\alpha}$ and $\beta$ represent the $4\times4$ usual Dirac matrices which are expressed in terms of the three $2\times2$ Pauli matrices and the $2\times2$ unit matrix. For spherical nuclei, the Dirac Hamiltonian commutes with the total angular momentum operator $\vec{J}=\vec{L}+\vec{S}$ and the spin-orbit coupling operator $\vec{K}=-\beta\left(\vec{\sigma}\cdot\vec{L}+1\right)$ where $\vec{L}$ and $\vec{S}$ are the orbital and spin momentum, respectively. The eigenvalues of the spin-orbit coupling operator are $\kappa=l>0$ and $\kappa=-(l+1)<0$ for unaligned spin $(j=l-1/2)$ and aligned $(j=l+1/2)$, respectively. Thus, the Dirac wave function takes the form:

$$\psi_{n\kappa}(\vec{r})=\frac{1}{r}\begin{pmatrix}F_{n\kappa}(r)Y_{jm}^{l}(\theta,\phi)\\ iG_{n\kappa}(r)Y_{jm}^{\tilde{l}}(\theta,\phi)\end{pmatrix}, \tag{19}$$

where $Y_{jm}^{l}(\theta,\phi)$ and $Y_{jm}^{\tilde{l}}(\theta,\phi)$ are the spin and p-spin spherical harmonics, respectively. $F_{n\kappa}(r)$ and $G_{n\kappa}(r)$ are the upper- and lower-spinor radial functions, respectively. Inserting Eq. (19) into Eq. (18) and using the relations [46]

$$\left(\vec{\sigma}\cdot\vec{A}\right)\left(\vec{\sigma}\cdot\vec{B}\right)=\vec{A}\cdot\vec{B}+i\vec{\sigma}\cdot\left(\vec{A}\times\vec{B}\right), \tag{20a}$$

$$\vec{\sigma}\cdot\vec{p}=\vec{\sigma}\cdot\hat{r}\left(\hat{r}\cdot\vec{p}+i\frac{\vec{\sigma}\cdot\vec{L}}{r}\right), \tag{20b}$$

and properties

$$\left(\vec{\sigma}\cdot\vec{L}\right)Y_{jm}^{\tilde{l}}(\theta,\phi)=(\kappa-1)Y_{jm}^{\tilde{l}}(\theta,\phi), \tag{21a}$$

$$\left(\vec{\sigma}\cdot\vec{L}\right)Y_{jm}^{l}(\theta,\phi)=-(\kappa+1)Y_{jm}^{l}(\theta,\phi), \tag{21b}$$

$$\left(\frac{\sigma\cdot r}{r}\right)Y_{jm}^{\tilde{l}}(\theta,\phi)=-Y_{jm}^{l}(\theta,\phi), \tag{21c}$$

$$\left(\frac{\sigma\cdot r}{r}\right)Y_{jm}^{l}(\theta,\phi)=-Y_{jm}^{\tilde{l}}(\theta,\phi), \tag{21d}$$

then we obtain the following two coupled differential equations satisfying the upper and lower radial functions $F_{n\kappa}(r)$ and $G_{n\kappa}(r)$ as

$$\left(\frac{d}{dr}+\frac{\kappa}{r}-U(r)\right)F_{n\kappa}(r)=\left(m+E-\Delta(r)\right)G_{n\kappa}(r), \tag{22a}$$

$$\left(\frac{d}{dr}-\frac{\kappa}{r}+U(r)\right)G_{n\kappa}(r)=\left(m-E+\Sigma(r)\right)F_{n\kappa}(r), \tag{22b}$$

where the sum and difference potentials are defined by

$$\Sigma(r)=V(r)+S(r) \text{ and } \Delta(r)=V(r)-S(r), \tag{23}$$

respectively. Combining Eqs. (22a) and (22b), we obtain the second-order differential equations satisfying the radial functions $F_{n\kappa}(r)$ and $G_{n\kappa}(r)$, respectively [47]

$$\left\{\frac{d^2}{dr^2}-\frac{\kappa(\kappa+1)}{r^2}+\left(\frac{2\kappa}{r}-U(r)-\frac{d}{dr}\right)U(r)-\left[m+E_{n\kappa}-\Delta(r)\right]\left[m-E_{n\kappa}+\Sigma(r)\right]\right.$$

$$\left.+\frac{1}{m+E_{n\kappa}-\Delta(r)}\frac{d\Delta(r)}{dr}\left(\frac{d}{dr}+\frac{\kappa}{r}-U(r)\right)\right\}F_{n\kappa}(r)=0, \tag{24}$$

$$\left\{\frac{d^2}{dr^2}-\frac{\kappa(\kappa-1)}{r^2}+\left(\frac{2\kappa}{r}-U(r)+\frac{d}{dr}\right)U(r)-\left[\left(m+E_{n\kappa}-\Delta(r)\right)\left(m-E_{n\kappa}+\Sigma(r)\right)\right]\right.$$

$$\left.-\frac{1}{m-E_{n\kappa}+\Sigma(r)}\frac{d\Sigma(r)}{dr}\left(\frac{d}{dr}-\frac{\kappa}{r}+U(r)\right)\right\}G_{n\kappa}(r)=0, \tag{25}$$

where $\kappa(\kappa+1)=l(l+1)$ and $\kappa(\kappa-1)=\tilde{l}\left(\tilde{l}+1\right)$. These radial wave equations are required to satisfy the necessary boundary conditions in the interval $r\in(0,\infty)$, i.e., $F_{n\kappa}(r=0)=G_{n\kappa}(0)=0$ and $F_{n\kappa}(r\to\infty)\simeq 0$, $G_{n\kappa}(r\to\infty)\simeq 0$.

### 3.1. Spin Symmetric Limit

Under the exact spin symmetric condition, we take the sum potential, $\Sigma(r)$, as the RM potential model, the difference potential, $\Delta(r)$, as a constant and the tensor potential, $U(r)$, as CLT potential. Then we have the following forms

$$\Sigma(r)=-4V_1\frac{e^{-2\alpha r}}{\left(1+e^{-2\alpha r}\right)^2}+V_2\frac{\left(1-e^{-2\alpha r}\right)}{\left(1+e^{-2\alpha r}\right)},\ \Delta(r)=C_s,\ \ U(r)=-\frac{H}{r}, \tag{26}$$

where $C_s$ and $H=Ze^2/\left(4\pi\varepsilon_0\right)$ are two constants.

Inserting Eq. (26) and the approximation given in [43,44] into Eq. (24)

[Dear Sameer

Here we must add the approximation as you used in J Math Phys]

and introducing a new parameter change $z(r) = -e^{-2\alpha r}$, this allows us to decompose the spin-symmetric Dirac equation (24) into the Schrödinger-like equation in the spherical coordinates for the upper-spinor component $F_{n\kappa}(r)$,

$$\left[\frac{d^2}{dz^2} + \frac{(1-z)}{z(1-z)}\frac{d}{dz} + \frac{\left(-\beta_1 z^2 + \beta_2 z - \varepsilon_{n\kappa}^2\right)}{z^2(1-z)^2}\right]F_{n\kappa}(z) = 0,\ F_{n\kappa}(z) = F_{n\kappa}(z) = 0, \tag{27a}$$

$$\varepsilon_{n\kappa} = \frac{1}{2\alpha}\sqrt{\frac{\omega}{r_e^2}D_0 - \tilde{E}_{n\kappa}^2 + \tilde{V}_2} > 0, \tag{27b}$$

$$\beta_1 = \frac{1}{4\alpha^2}\left[\frac{\omega}{r_e^2}\left(D_0 + D_1 + D_2\right) - \tilde{E}_{n\kappa}^2 - \tilde{V}_2\right], \tag{27c}$$

$$\beta_2 = \frac{1}{4\alpha^2}\left[\frac{\omega}{r_e^2}\left(2D_0 + D_1\right) - 2\tilde{E}_{n\kappa}^2 - 4\tilde{V}_1\right], \tag{27d}$$

where $\tilde{V}_1 = V_1\left(m + E_{n\kappa} - C_s\right)$, $\tilde{V}_2 = V_2\left(m + E_{n\kappa} - C_s\right)$, $\tilde{E}_{n\kappa}^2 = \left(E_{n\kappa} + m - C_s\right)\left(E_{n\kappa} - m\right)$ and $\omega = \left(\kappa + H\right)\left(\kappa + H + 1\right)$. Further, the explicit forms of the constants $D_i\ (i = 1,2,3)$ are defined in [43,44] and are being expressed in terms of the potential parameters. In order to solve Eq. (27a) by means of the NU method, we should compare it with Eq. (2) to obtain the following particular values for the parameters:

$$\tilde{\tau}(z) = 1 - z,\ \sigma(z) = z(1-z),\ \tilde{\sigma}(z) = -\beta_1 z^2 + \beta_2 z - \varepsilon_{n\kappa}^2. \tag{28}$$

Comparing Eq. (29) with Eq. (5), we can easily obtain the coefficients $c_i\ (i = 1,2,3,4)$ and the analytical expressions $A$, $B$ and $C$. However, the values of the coefficients $c_i\ (i = 5,6,\cdots,16)$ are found from the relations (6). Therefore, the specific values of the coefficients $c_i\ (i = 1,2,\cdots,16)$ together with $A$, $B$ and $C$ are

$$c_1 = c_2 = c_3 = c_4 = 1,\ c_5 = 0,\ c_6 = -\frac{1}{2},\ c_7 = \frac{1}{4} + \beta_1,\ c_8 = -\beta_2,\ c_9 = \varepsilon_{n\kappa}^2,\ c_{10} = \left(\delta + \frac{1}{2}\right)^2,$$

$$c_{11} = 2\varepsilon_{n\kappa},\ c_{12} = c_{15} = 2\delta + 1,\ c_{13} = \varepsilon_{n\kappa},\ c_{14} = c_{16} = \delta + 1,\ A = \beta_1,\ B = \beta_2,\ C = \varepsilon_{n\kappa}^2,$$

$$\delta = \frac{1}{2}\left(-1 + \sqrt{1 + \frac{1}{\alpha^2}\left(\frac{\omega D_2}{r_e^2} + 4\tilde{V}_1\right)}\right). \tag{29}$$

From Eqs. (7)-(9) together with the coefficients given in Eq. (29), we can calculate the essential parameters $\pi(z)$, $k$ and $\tau(z)$ as

$$\pi(z) = \varepsilon_{n\kappa} - \left(1 + \varepsilon_{n\kappa} + \delta\right)z, \tag{30}$$

$$k=\beta_2-\left[2\varepsilon_{n\kappa}^2+(2\delta+1)\varepsilon_{n\kappa}\right], \tag{31}$$

and

$$\tau(z)=1+2\varepsilon_{n\kappa}-(3+2\varepsilon_{n\kappa}+2\delta)z,\quad \tau'(z)=-(3+2\varepsilon_{n\kappa}+2\delta)<0, \tag{32}$$

with prime denotes the derivative with respect to $z$. Equation (11) gives the energy equation for the RM potential including the CLT potential in the Dirac theory as

$$(m+E_{n\kappa}-C_s)(m-E_{n\kappa}+V_2)=-\frac{\omega D_0}{r_e^2}$$
$$+\alpha^2\left[\frac{-2V_1(m+E_{n\kappa}-C_s)+\omega(D_1+D_2)/r_e^2}{4\alpha^2(n+\delta+1)}-(n+\delta+1)\right]^2. \tag{33}$$

Further, for the exact spin symmetric case, $S(\vec{r})=V(\vec{r})$ or $C_s=0,$ we obtain

$$(m+E_{n\kappa})(m-E_{n\kappa}+V_2)=-\frac{\omega D_0}{r_e^2}$$
$$+\alpha^2\left[\frac{-2V_1(m+E_{n\kappa})+\omega(D_1+D_2)/r_e^2}{4\alpha^2(n+\tilde{\delta}+1)}-(n+\tilde{\delta}+1)\right]^2,$$

with $\tilde{\delta}=\delta(C_s\to 0)=\frac{1}{2}\left(-1+\sqrt{1+\frac{1}{\alpha^2}\left(\frac{\omega D_2}{r_e^2}+4V_1(m+E_{n\kappa})\right)}\right).$

Let us now find the corresponding wave functions for this potential model. Referring to Eqs. (12), (13) and (29), we find the functions:

$$\rho(z)=z^{2\varepsilon_{n\kappa}}(1-z)^{2\delta+1}, \tag{34}$$

$$\phi(z)=z^{\varepsilon_{n\kappa}}(1-z)^{\delta+1},\ \varepsilon_{n\kappa}>0,\ \delta>0, \tag{35}$$

Hence, Eq. (14) gives

$$y_n(z)=A_n z^{-2\varepsilon_{n\kappa}}(1-z)^{-(2\delta+1)}\frac{d^n}{dz^n}\left[z^{n+2\varepsilon_{n\kappa}}(1-z)^{n+2\delta+1}\right]$$
$$=P_n^{(2\varepsilon_{n\kappa},2\delta+1)}(1-2z),\ 2\varepsilon_{n\kappa}>-1,\ 2\delta+1>-1,\ z\in[0,1], \tag{36}$$

where the Jacobi polynomials $P_n^{(\mu,\nu)}(x)$, where $\mu>-1,$ $\nu>-1$ and $x\in[-1,+1]$. By using $F_{n\kappa}(z)=\phi(z)y_n(z),$ we get the radial upper-spinor wave functions from Eq. (15) as

$$F_{n\kappa}(r)=N_{n\kappa}\left(-e^{-2\alpha r}\right)^{\varepsilon_{n\kappa}}\left(1+e^{-2\alpha r}\right)^{\delta+1}P_n^{(2\varepsilon_{n\kappa},2\delta+1)}\left(1+2e^{-2\alpha r}\right)$$
$$=N_{n\kappa}\left(e^{-2\alpha r}\right)^{\varepsilon_{n\kappa}}\left(1+e^{-2\alpha r}\right)^{\delta+1}{}_2F_1\left(-n,n+2(\varepsilon_{n\kappa}+\delta+1);2\varepsilon_{n\kappa}+1;-e^{-2\alpha r}\right), \tag{37}$$

where the normalization constant has been calculated in Ref. [44]. The lower component $G_{n\kappa}(r)$ can be obtained as follows [44]

$$G_{n\kappa}(r)=N_{n\kappa}\frac{\left(-e^{-2\alpha r}\right)^{\varepsilon_{n\kappa}}\left(1+e^{-2\alpha r}\right)^{\delta+1}}{\left(m+E_{n\kappa}-C_s\right)}\left[-2\alpha\varepsilon_{n\kappa}-\frac{2\alpha(\delta+1)e^{-2\alpha r}}{\left(1+e^{-2\alpha r}\right)}+\frac{\kappa}{r}\right]$$

$$\times{}_2F_1\left(-n,n+2(\varepsilon_{n\kappa}+\delta+1);2\varepsilon_{n\kappa}+1;-e^{-2\alpha r}\right)$$

$$+N_{n\kappa}\left[\frac{2\alpha n\left[n+2\left(\varepsilon_{n\kappa}+\delta+1\right)\right]\left(-e^{-2\alpha r}\right)^{\varepsilon_{n\kappa}+1}(1+e^{-2\alpha r})^{\delta+1}}{\left(2\varepsilon_{n\kappa}+1\right)\left(m+E_{n\kappa}-C_s\right)}\right]$$

$$\times{}_2F_1\left(-n+1,n+2(\varepsilon_{n\kappa}+\delta+3/2);2\left(\varepsilon_{n\kappa}+1\right);-e^{-2\alpha r}\right), \tag{38}$$

where $E_{n\kappa}\neq -m$ for exact spin symmetry. Hence, the spin symmetric solution has only positive energy spectrum. Here, it should be noted that the hypergeometric series ${}_2F_1\left(-n,n+2(\varepsilon_{n\kappa}+\delta+1);2\varepsilon_{n\kappa}+1;-e^{-2\alpha r}\right)$ terminates for $n=0$ state and thus does not diverge for all values of real parameters $\delta$ and $\varepsilon_{n\kappa}$.

### 3.2. P-Spin Symmetric Limit

Inserting $\Delta(r)$ as the RM potential, $\Sigma(r)$ as a constant and $U(r)$ as Coulomb-like tensor potential, i.e.,

$$\Delta(r)=-4V_1\frac{e^{-2\alpha r}}{\left(1+e^{-2\alpha r}\right)^2}+V_2\frac{\left(1-e^{-2\alpha r}\right)}{\left(1+e^{-2\alpha r}\right)},\ \Sigma(r)=C_{ps},\ \ U(r)=-\frac{H}{r}, \tag{39}$$

into Eq. (25), we obtain the following Schrödinger-like equation for the lower-spinor component $G_{n\kappa}(r)$,

$$\left[\frac{d^2}{dz^2}+\frac{(1-z)}{z(1-z)}\frac{d}{dz}+\frac{\left(-\bar{\beta}_1z^2+\bar{\beta}_2z-\bar{\varepsilon}_{n\kappa}^2\right)}{z^2(1-z)^2}\right]G_{n\kappa}(z)=0, \tag{40a}$$

$$\bar{\varepsilon}_{n\kappa}=\frac{1}{2\alpha}\sqrt{\frac{\bar{\omega}}{r_e^2}D_0-\bar{E}_{n\kappa}^2+\bar{V}_2}>0, \tag{41b}$$

$$\bar{\beta}_1=\frac{1}{4\alpha^2}\left[\frac{\bar{\omega}}{r_e^2}\left(D_0+D_1+D_2\right)-\bar{E}_{n\kappa}^2-\bar{V}_2\right], \tag{41c}$$

$$\bar{\beta}_2=\frac{1}{4\alpha^2}\left[\frac{\bar{\omega}}{r_e^2}\left(2D_0+D_1\right)-2\bar{E}_{n\kappa}^2-4\bar{V}_1\right], \tag{41d}$$

where $\bar{V}_1 = V_1\left(E_{n\kappa} - m - C_{ps}\right)$, $\bar{V}_2 = V_2\left(E_{n\kappa} - m - C_{ps}\right)$, $\bar{E}_{n\kappa}^2 = \left(E_{n\kappa} - m - C_{ps}\right)\left(E_{n\kappa} + m\right)$ and $\bar{\omega} = (\kappa + H)(\kappa + H - 1)$.

To avoid repetition in the solution of Eq. (41a), a first inspection for the relationship between the present set of parameters $(\bar{\varepsilon}_{n\kappa}, \bar{\beta}_1, \bar{\beta}_2)$ and the previous set $(\varepsilon_{n\kappa}, \beta_1, \beta_2)$ tells us that the negative energy solution for p-spin symmetry, where $S(\vec{r}) = -V(\vec{r})$, can be obtained directly from those of the above positive energy solution for spin symmetry by using the parameter mapping [44]:

$$F_{n\kappa}(r) \leftrightarrow G_{n\kappa}(r);\ V(r) \to -V(r)\ (V_1 \to -V_1, V_2 \to -V_2);\ E_{n\kappa} \to -E_{n\kappa};\ C_s \leftrightarrow -C_{ps}. \tag{42}$$

Following the previous results with the above transformations, we finally arrive at the energy equation

$$\left(m - E_{n\kappa} + C_{ps}\right)\left(m + E_{n\kappa} - V_2\right) = -\frac{\bar{\omega} D_0}{r_e^2} + \alpha^2 \left[ \frac{2V_1\left(m - E_{n\kappa} + C_{ps}\right) + \bar{\omega}\left(D_1 + D_2\right)/r_e^2}{4\alpha^2\left(n + \eta + 1\right)} - \left(n + \eta + 1\right) \right]^2, \tag{43}$$

with

$$\eta = \frac{1}{2}\left( -1 + \sqrt{1 + \frac{1}{\alpha^2}\left( \frac{\bar{\omega} D_2}{r_e^2} + 4\bar{V}_1 \right)} \right). \tag{44}$$

By using $G_{n\kappa}(r) = \phi(z) y_n(z)$, we get the radial lower-spinor wave functions as

$$G_{n\kappa}(r) = \tilde{N}_{n\kappa}\left(-e^{-2\alpha r}\right)^{\bar{\varepsilon}_{n\kappa}}\left(1 + e^{-2\alpha r}\right)^{\eta+1} P_n^{(2\bar{\varepsilon}_{n\kappa}, 2\eta+1)}\left(1 + 2e^{-2\alpha r}\right), \tag{45}$$

where $G_{n\kappa}(r)$ satisfies the restriction condition for the bound states, i.e., $\eta > 0$ and $\bar{\varepsilon}_{n\kappa} > 0$. The normalization constants $\tilde{N}_{n\kappa}$ are calculated in Ref. [44].

To check our analytical expressions, we have calculated the energy levels for the p-spin and spin cases in Tables 1 and 2, respectively using the following set of parameter values: $m = 1.0\,fm^{-1}$, $V_1 = 1.0\,fm^{-1}$, $V_2 = -1.0\,fm^{-1}$ $r_e = 2.197224577\,fm$, $\alpha = 0.25\,fm^{-1}$, $C_s = 0\,fm^{-1}$ and $C_{ps} = -6.0\,fm^{-1}$. In table 1, we observe the degeneracy in the following doublets $\left(1s_{1/2}, 0d_{3/2}\right), \left(1p_{3/2}, 0f_{5/2}\right), \left(1d_{5/2}, 0g_{7/2}\right)$, $\left(1f_{7/2}, 0h_{9/2}\right)$, and so on. Thus, each pair is considered as p-spin doublet and has negative energy. In table 2, we present the energy spectrum for the spin symmetric case. Obviously, the pairs $\left(np_{1/2}, np_{3/2}\right), \left(nd_{3/2}, nd_{5/2}\right), \left(nf_{5/2}, nf_{7/2}\right), \left(ng_{7/2}, ng_{9/2}\right)$, and so forth are degenerate states. Thus, each pair is considered as spin doublet and has positive energy. The numerical results in Tables 1 and 2 emphasize that the presence of the CLT interaction removes degeneracies between spin and p-spin state doublets.

Finally, we plot the relativistic energy eigenvalues of the RM potential and CLT potential with spin and p-spin symmetry limitations in Figures 1 to 4. Figures 1 and 2 show the variation of the energy levels versus the Coulomb tensor strength $H$ and the screening parameter $\alpha$ respectively in the case of p-spin symmetry considering the following pairs of orbital states $\left(1d_{5/2}, 0g_{7/2}\right)$, $\left(2f_{7/2}, 1h_{9/2}\right)$, $\left(3g_{9/2}, 2i_{11/2}\right)$. From Fig. 1, we observe that in the case of $H = 0$ (no tensor interaction), members of p-spin doublets have same energy. However, in the presence of the tensor potential $H \neq 0$, these degeneracies are removed. We can also see in Fig. 1 that p-spin doublet splitting increases with increasing $H$. The reason is that term $2\kappa H$ gives different contributions to each level in the spin doublet because $H$ takes different values for each state in the spin doublet. In Fig. 2, the contribution of the screening parameter $\alpha$ to the p-spin doublet splitting is presented. It can be seen that magnitude of the energy difference between members of the p-spin doublet decreases as $\alpha$ increases.

Further, in Figures 3 and 4, we have investigated the effect of Coulomb tensor strength $H$ and the screening parameter $\alpha$ on the spin doublet splitting by considering the following orbital pairs: $\left(1p_{1/2}, 1p_{3/2}\right)$, $\left(1d_{3/2}, 1d_{5/2}\right)$ and $\left(1f_{5/2}, 1f_{7/2}\right)$ and one can observe that the results obtained in the spin symmetric limit resemble the ones observed in the p-spin symmetric limit.

## 4. Conclusion

We have obtained analytically the spin and p-spin symmetric energy eigenvalues and the corresponding wave functions of the Dirac-RM problem with CLT potential in the frame of the NU method. For any spin-orbit quantum number $\kappa$, we have found the approximate expressions for the energy eigenvalues and associated wave functions in closed form. The numerical results indicate that the CLT interaction removes degeneracies in spin and p-spin state doublets.

The most stringent interesting result is that the present spin and p-spin symmetric cases can be easily reduced to the KG solution once $S(\vec{r}) = V(\vec{r})$ and $S(\vec{r}) = -V(\vec{r})$ (i.e., $C_s = C_{ps} = 0$) [36]. The resulting solutions of the wave functions are being expressed in terms of the generalized Jacobi polynomials. Obviously, the relativistic solution can be reduced to it's non-relativistic limit by the choice of appropriate mapping transformations [44]. Also, in case when spin-orbit quantum number $\kappa = 0$, the problem reduces to the $s$ -wave solution.

**Acknowledgments**

S. M. Ikhdair acknowledges the partial support provided by the Scientific and Technological Research Council of Turkey (TÜBİTAK).

**Table 1.** The energy levels in units of $fm^{-1}$ of the p-spin symmetry RM potential for several values of $n$ and $\kappa$ with $H=0.5$.

| $\tilde{l}$ | $n,\kappa<0$ | $(l,j)$ | $E_{n,\kappa<0}$ $H\neq 0$ | $E_{n,\kappa<0}$ $H=0$ | $n-1,\kappa>0$ | $(l+2,j+1)$ | $E_{n-1,\kappa>0}$ $H\neq 0$ | $E_{n-1,\kappa>0}$ $H=0$ |
|---|---|---|---|---|---|---|---|---|
| 1 | 1, -1 | $1s_{1/2}$ | −1.903134794 | −1.738772757 | 0, 2 | $0d_{3/2}$ | −1.538608678 | −1.738772757 |
| 2 | 1, -2 | $1p_{3/2}$ | −1.538608678 | −1.313563183 | 0, 3 | $0f_{5/2}$ | −1.071543282 | −1.313563183 |
| 3 | 1, -3 | $1d_{5/2}$ | −1.071543282 | −.8177171059 | 0, 4 | $0g_{7/2}$ | −.5554514514 | −.8177171059 |
| 4 | 1, -4 | $1f_{7/2}$ | −.5554514514 | −.2869876340 | 0, 5 | $0h_{9/2}$ | −0.01385847616 | −.2869876340 |
| 1 | 2, -1 | $2s_{1/2}$ | −1.921760586 | −1.745387730 | 1, 2 | $1d_{3/2}$ | −1.521903111 | −1.745387730 |
| 2 | 2, -2 | $2p_{3/2}$ | −1.521903111 | −1.273143731 | 1, 3 | $1f_{5/2}$ | −1.01005856 | −1.273143731 |
| 3 | 2, -3 | $2d_{5/2}$ | −1.010058564 | −.7382341146 | 1, 4 | $1g_{7/2}$ | −.4607283791 | −.7382341146 |
| 4 | 2, -4 | $2f_{7/2}$ | −.4607283791 | −.1793345097 | 1, 5 | $1h_{9/2}$ | 0.1048326489 | −.1793345097 |

**Table 2. The** energy levels in units of $fm^{-1}$ of the spin symmetry RM potential for several values of $n$ and $\kappa$ with $H=0.5$.

| $l$ | $n,\kappa<0$ | $(l, j=l+1/2)$ | $E_{n,\kappa<0}$ $H\neq 0$ | $E_{n,\kappa<0}$ $H=0$ | $n,\kappa>0$ | $(l, j=l-1/2)$ | $E_{n,\kappa>0}$ $H\neq 0$ | $E_{n,\kappa>0}$ $H=0$ |
|---|---|---|---|---|---|---|---|---|
| 1 | 0, -2 | $0p_{3/2}$ | 0.1483955852 | 0.3935828782 | 0, 1 | $0p_{1/2}$ | 0.6582373104 | 0.3935828782 |
| 2 | 0, -3 | $0d_{5/2}$ | 0.6582373104 | 0.9333946490 | 0, 2 | $0d_{3/2}$ | 1.214682872 | 0.9333946490 |
| 3 | 0, -4 | $0f_{7/2}$ | 1.214682872 | 1.499799551 | 0, 3 | $0f_{5/2}$ | 1.787442858 | 1.499799551 |
| 4 | 0, -5 | $0g_{9/2}$ | 1.787442858 | 2.076830625 | 0, 4 | $0g_{7/2}$ | 2.367468347 | 2.076830625 |
| 1 | 1, -2 | $1p_{3/2}$ | 0.2663841239 | 0.5510806494 | 1, 1 | $1p_{1/2}$ | 0.8381866907 | 0.5510806494 |
| 2 | 1, -3 | $1d_{5/2}$ | 0.8381866907 | 1.127479271 | 1, 2 | $1d_{3/2}$ | 1.418299684 | 1.127479271 |
| 3 | 1, -4 | $1f_{7/2}$ | 1.418299684 | 1.710188317 | 1, 3 | $1f_{5/2}$ | 2.002843683 | 1.710188317 |
| 4 | 1, -5 | $1g_{9/2}$ | 2.002843683 | 2.296064965 | 1, 4 | $1g_{7/2}$ | 2.589714652 | 2.296064965 |

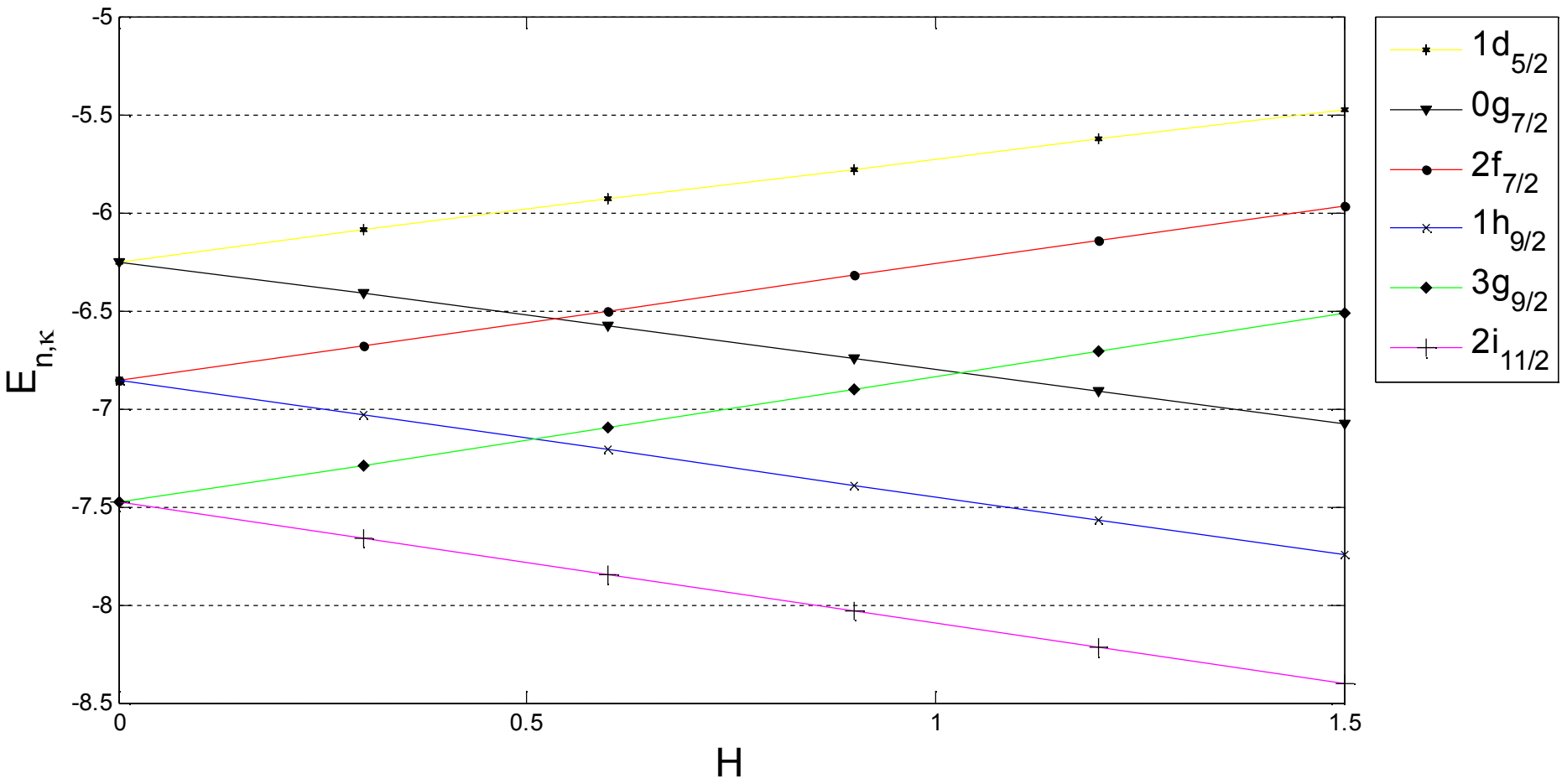

**Figure 1.** The energy levels versus the CLT strength $H$ in the p-spin symmetric case using the parameters. $m = 1.0\,fm^{-1}$, $C_{ps} = -6.0\,fm^{-1}$, $\alpha = 0.25\,fm^{-1}$, $r_e = 2.197224577\,fm$, $V_1 = 1.0\,fm^{-1}$ and $V_2 = -1.0\,fm^{-1}$.

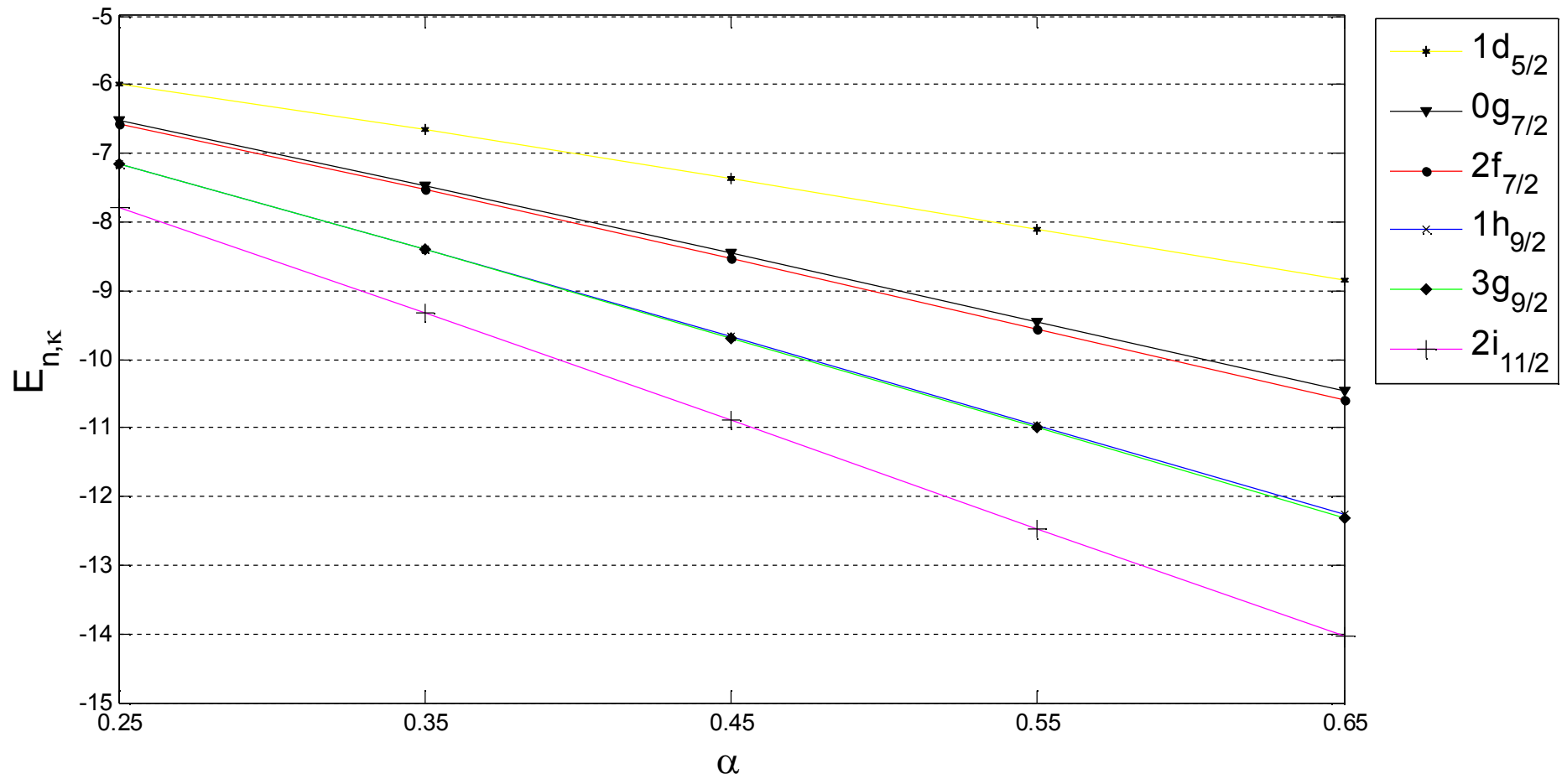

**Figure 2.** The energy levels versus the screening parameter $\alpha$ in the p-spin symmetric case using the parameters. $m = 1.0\,fm^{-1}$, $C_{ps} = -6.0\,fm^{-1}$, $H = 0.5$, $r_e = 2.197224577\,fm$, $V_1 = 1.0\,fm^{-1}$ and $V_2 = -1.0\,fm^{-1}$.

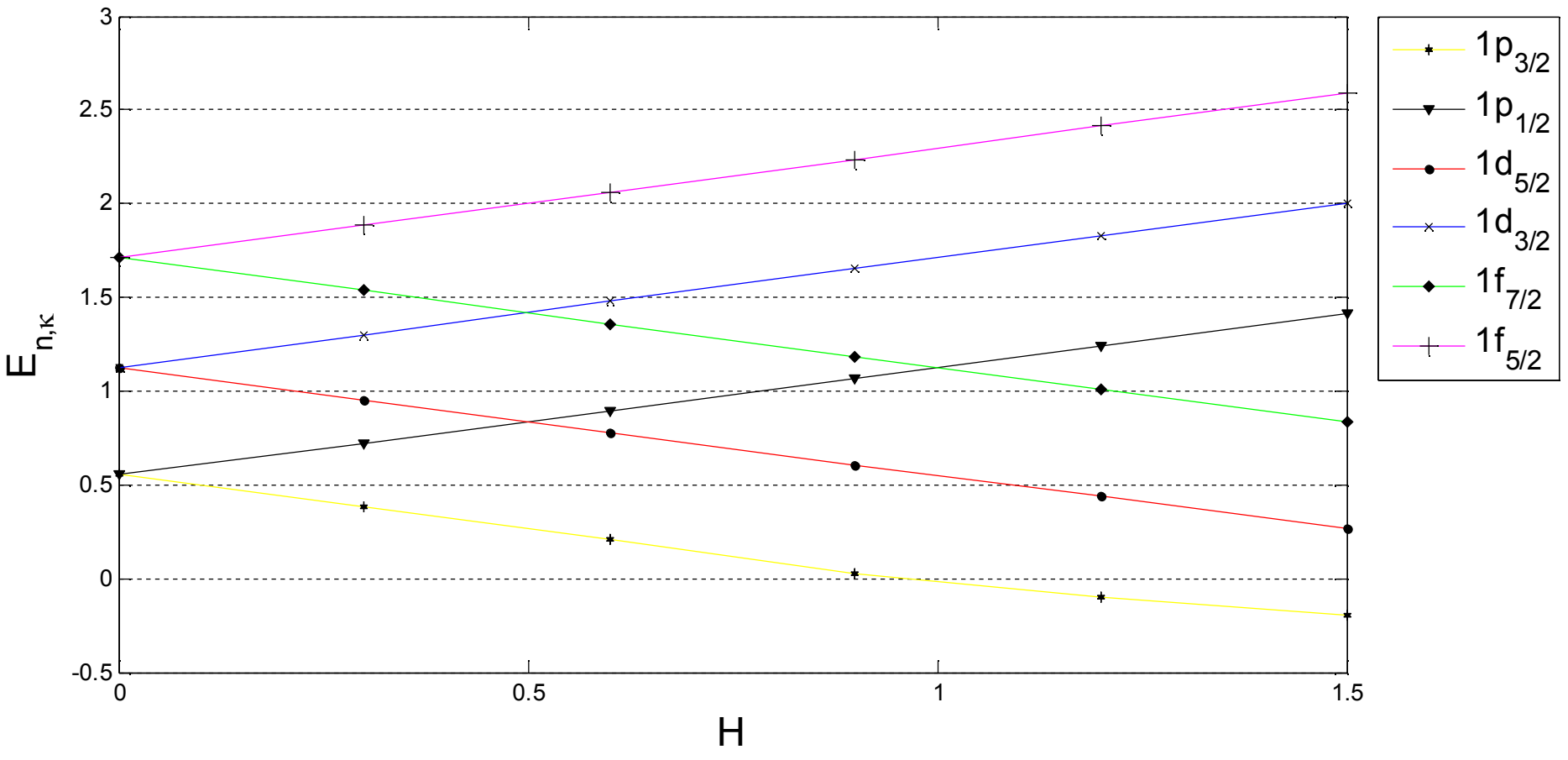

**Figure 3.** The energy levels versus the CLT strength $H$ in the spin symmetric case using the parameters. $m = 1.0\,fm^{-1}$, $C_s = 0\,fm^{-1}$, $\alpha = 0.25\,fm^{-1}$, $r_e = 2.197224577\,fm$, $V_1 = 1.0\,fm^{-1}$ and $V_2 = -1.0\,fm^{-1}$.

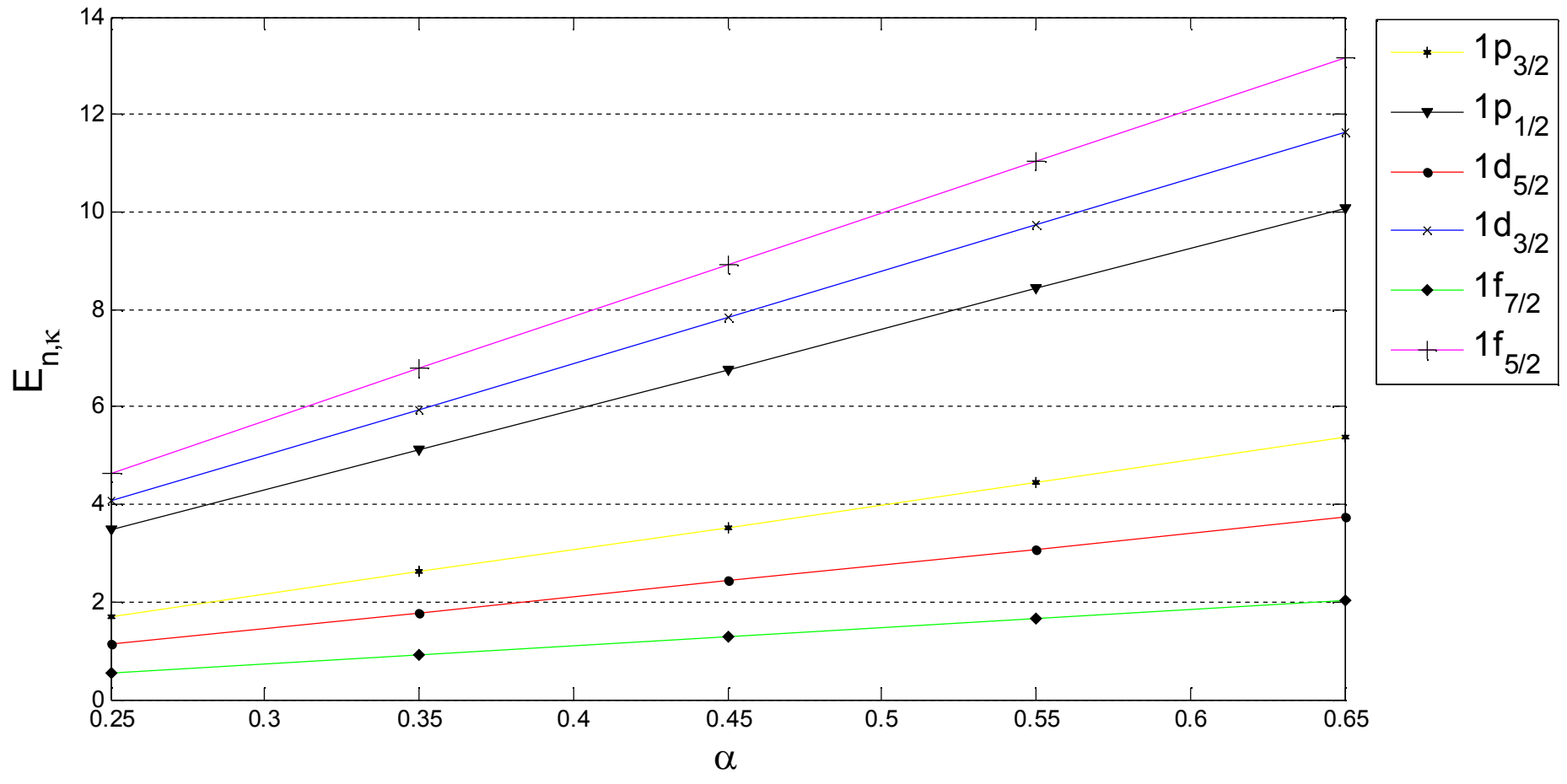

**Figure 4.** The energy levels versus the screening parameter $\alpha$ in the spin symmetric case using the parameters. $m = 1.0\,fm^{-1}$, $C_s = 0\,fm^{-1}$, $H = 5.0$, $r_e = 2.197224577\,fm$, $V_1 = 1.0\,fm^{-1}$ and $V_2 = -1.0\,fm^{-1}$.